# From Research to Resources: Assessing Student Understanding and Skills in Quantum Computing

Beth Thacker[1], Jianlan Wang[1], Yuanlin Zhang[1], Quy Ban Tran[1] Divya Sree Vemula[1], and Tunde Kushimo[2]

[1]Texas Tech University, [2]Wichita State University

*Contact information for lead author: Beth Thacker, Department of Physics and Astronomy, Texas Tech University, Lubbock, TX, 79416, email:beth.thacker@ttu.edu; phone: 806-834-2996*

**Abstract:** The revolutionary new field of Quantum Computing (QC) continues to gain attention in industry, academia, and government in both research and education. At educational institutions, there is a proliferation of introductory courses at various academic levels signaling a growing interest and recognition of the significance of this field. A crucial and often overlooked aspect is the development of research-based materials and pedagogical approaches to effectively teach the complexities of quantum computing to diverse cohorts of learners across multiple disciplines. There is a great need for empirical investigations of the effectiveness of learning materials and pedagogical approaches in this new interdisciplinary field. We present an empirical investigation done at an R1 institution using the multiple case study method. We compare a case study on students in an introductory QC course without research-based mini-tutorials to a study of students taking the QC course with research-based mini-tutorials. We compare the strengths and difficulties of students in the two courses, discuss the general strengths and difficulties of students across both courses, postulate the effectiveness of the mini-tutorials and discuss how they can be revised. Strengths across both classes include the ability to apply single-qubit and two-qubit gates, favoring application of Dirac notation, and a reasonable understanding of normalization, probability and teleportation described qualitatively. Difficulties across both classes included use of matrix representation, use of rotation gates, and an ability to recall and analyze quantitatively a circuit representing teleportation. We postulate that the mini-tutorials on gates, normalization, probability and a qualitative description of teleportation may have been effective in increasing students' understanding of the concepts. We also identify mini-tutorials that may need revision.

# I. INTRODUCTION

The revolutionary new field of Quantum Computing (QC) continues to gain attention in industry, academia, and government in both research and education. At educational institutions, there is a proliferation of introductory courses at various academic levels signaling a growing interest and recognition of the significance of this field [1]. A crucial aspect often overlooked is the development of research-based materials and pedagogical approaches to effectively teach the complexities of quantum computing to diverse cohorts of learners across multiple disciplines. While this research has begun [2-12], there is still a great need for further empirical investigations of 1) students' QC conceptual understanding as evidenced by their application of those concepts in analyses of QC circuits and protocols and 2) the effectiveness of learning materials and pedagogical approaches in this new interdisciplinary field.

While there has been significant study of students' understanding of quantum mechanics concepts in the context of quantum mechanics courses at the undergraduate and graduate levels [13-15], the research on students' understanding of QC concepts and skills at applying those concepts in the context of a QC course is not that extensive [8-12]. There has been some research on students' understanding of QC concepts in the context of quantum mechanics courses [2-7], but those courses are usually designed for upper-level or graduate physics majors and do not focus explicitly on QC concepts in the context of an interdisciplinary QC course. Our investigation adds to the existing research on students' understanding of QC concepts and skills at applying those concepts in the context of QC courses.

In this paper, we present an empirical investigation done at an R1 University using the multiple case study method [16]. We compare a case study on students in an inaugural introductory QC course in Fall 2021 without research-based mini-tutorials (F21) to a case study of students taking the QC course two years later in Fall 2023 with research-based mini-tutorials (F23). The mini-tutorials are patterned after physics tutorials developed by various physics education research groups, such as the University of Washington [17]. However, they are shorter, designed to be done in 20-30 minutes in class, and focus on enhancing the ability to apply important concepts and skills covered in class. The research methodology involves conducting interviews with students and analyzing the interview transcripts to identify students' strengths and difficulties in understanding QC concepts and the application of those concepts in theoretical analyses of quantum computing circuits and protocols. We also compared the two cases of students taught with and without the mini-tutorials.

Our research questions are:

1. What are the strengths and difficulties of students in an introductory QC course in understanding QC concepts as evidenced in the application of those concepts in theoretical analyses of quantum computing circuits and protocols?
2. How do the strengths and difficulties of students in understanding QC concepts in an introductory quantum computing class taught with and without research-based mini-tutorials compare?
3. Do we have evidence that differences in students' strengths and difficulties in an introductory quantum computing class taught with and without research-based mini-tutorials are due to the use of the mini-tutorials?

In previous work [12], we identified students' strengths and difficulties in an introductory QC class (F21), as evidenced by their understanding of important concepts and their skill at applying those concepts in quantum computing circuits and protocols. In that qualitative study, we used semi-structured interviews with student volunteers and analyzed the interviews using the thematic analysis method. We used the results of the study as the basis for the development of a set of mini-tutorials focused on addressing students' difficulties in learning QC concepts and their skills in applying those concepts.

In this study, we interviewed students in the F23 section of the course with a subset of the same interview questions used in the F21 course interviews. In each case, the interviews were done the semester after the course was taught, Spring 2022 (S22) and Spring 2024 (S24), respectively. We also developed a rubric to analyze each set of interviews with a particular focus on the objectives of the mini-tutorials. We employed the rubric on both sets of data (re-analyzing the first set of data with the new rubric). Our mini-tutorial development, rubric development, and interview analysis was done through a lens of constructivist learning theory [18-20]. The student-centered interactive engagement class was taught with learners as active participants, developing their mental models based on their previous knowledge and aided by their social in-class interactions. The mini-tutorials were designed to facilitate the application of QC concepts to theoretical analyses of QC circuits and protocols.

We report on the strengths and difficulties of students observed in the interviews in each case study and a comparison of both case studies.

## II. QUANTUM COMPUTING COURSE AND STUDENT POPULATION

### A. Fall 2021 Inaugural Course and Student Population

The course was first taught in F21. It was taught as an upper-level Physics Elective and there were 6 students in the course, all junior or senior physics majors. The first 2-3 weeks of the course focused on the physics of two-level quantum systems as taught in the Quantum Mechanics text by David McIntyre [21]. After that, the main text was the 2021 version of the online *Qiskit* textbook by IBM (now the legacy version) [22] supplemented by materials from other texts [23-35]. The course was taught interactively in a semi-flipped format, with students asked, but not required, to read ahead and a significant amount of class time spent on discussions, problems, and exercises. There were two exams, a midterm and a final, graded homework, a project, and a participation grade for attending class, contributing to discussions, group work on problems, exercises, etc. There were no pre-requisites for the course. The quantum mechanics and linear algebra needed were taught within the course. The topics covered and an approximate schedule are shown in Table 1.

### B. Subsequent Course Offerings, Student Populations, and Introduction of Mini-tutorials.

The course has been taught each Fall semester since the inaugural course. The instructor continued to use the MacIntyre physics text, the IBM Qiskit textbook and the same approximate schedule, coverage and format (semi-flipped, very interactive, a significant amount of class time spent on discussions, problems, and exercises). In the second and third semesters, there were 10 and 11 students in the course, respectively.

**TABLE 1**. Topics covered and approximate schedule.

| Week | Topic |
|---|---|
| 1 | Stern-Gerlach Experiment, Quantum State Vectors |
| 2 | Matrix notation, Operators, Eigenvalues and Eigenvectors, |
| 3 | Measurement, Commuting Observables, Uncertainty Principle |
| 4 | Classical Bits, Qubits, Single Qubit Gates |
| 5 | Big O notation, No-Cloning Theorem |
| 6 | Multiple Qubits and Entangled States |
| 7 | Phase Kickback, Circuit Identities |
| 8 | Proving Universality, Quantum Teleportation, Superdense Coding |
| 9 | Deutsch Algorithm, Deutsch-Jozsa Algorithm |
| 10 | Bernstein-Vazarani Algorithm |
| 11 | Simon's Algorithm |
| 12 | Quantum Fourier Transform |
| 13 | Quantum Phase Estimation |

In F23, the course was co-listed for the first time in Physics and Electrical and Computer Engineering (ECE). In F23, we introduced a set of mini-tutorials designed to focus on or expand on certain topics. The mini-tutorials are patterned after research-based physics tutorials developed by various physics education research groups, such as the University of Washington [20]. However, they are shorter, designed to be done in 20-30 minutes, and focus on enhancing the ability to apply important concepts and skills covered in class. The mini-tutorial is not the first time a student is exposed to a concept or skill. It is usually designed so that the student is required to apply the concept/skill in a different context than was initially covered in class. This will help the student know if they actually understand the concept/skill and are able to apply it. The development of the mini-tutorials, based on the previous research [12], was a mechanism for the instructor to formalize some of the classroom discussions, problems and exercises previously done in the class. The topics addressed in the mini-tutorials were:

- Probability in the context of a Stern-Gerlach analyzer
- Half Adder
- Single Qubit Gates
- Generating Bell States
- Entangled States
- Phase Kickback
- Teleportation
- Deutsch's Algorithm
- Deutsch-Josza Algorithm
- Quantum Fourier Transform

The intent of each mini-tutorial was different. As the F21 interviews had indicated that students had difficulty with applying different representations to circuit problems, some mini-tutorials were developed that required students to answer the same question using different representations. For example, in the Single Qubit Gates mini-tutorial, students were asked to answer the same question in three different representations: Matrix, Dirac, and Bloch Sphere. In other cases, the student was to solidify their knowledge by working through a protocol or an algorithm with different initial conditions than had been used in class. In the quantum teleportation mini-tutorial, for example, the students are first asked to describe teleportation qualitatively and then to work through the circuit starting with a different Bell state than had been used in class. The Deutsch, Deutsch-Josza and Quantum Teleportation algorithms also required students to work through the algorithms with different initial conditions. Sample mini-tutorials are shown in Appendix A.

### C. Interviewee Populations

In this paper, we discuss interviews with students in the first and third semesters of the course, F21 and F23.

As stated above, in F21, there were six physics majors in the course, four seniors, and two juniors. Although the course was listed without pre-requisites, all of the students had taken junior-level quantum mechanics or were taking it concurrently and had taken linear algebra. We report on interviews with four students the semester after they took the course, Spring 2022 (S22). Information on the students is listed in Table 2, including their education level, level based on exam grades, major, and whether or not they had taken quantum mechanics and linear algebra at the time of taking the course. The level based on exam grades can be interpreted as High Proficiency (A), Proficient (A-/B+), and Lower Proficiency (C) on exams. For example, F21_1_a refers to a High Proficiency student interviewee from F21. The letters "a" and "b" distinguish individuals.

In F23, when the course was co-listed for the first time in Physics and ECE, there were eleven students in the course, five undergraduate Computer Engineering majors, two undergraduate Physics majors, one undergraduate Electrical Engineering major, one undergraduate Economics major, one Electrical Engineering graduate student and one Interdisciplinary Studies graduate student. We report on interviews with four students the semester after they took the course (S24). All of the students who volunteered to be interviewed were Computer Engineering majors. The Computer Engineering majors interviewed had not had physics beyond Physics II, Electricity and Magnetism, so they had not studied Quantum Mechanics before taking the course. Some of them had taken Linear Algebra. Information on the students is listed in Table 2. We list whether or not the students had taken quantum mechanics or linear algebra before taking the QC course, but neither of these were pre-requisites for the course.

## III. METHODS

Clinical interviews [36] with the participants served as the data source for this study. To help identify strenghts and difficulties, we developed and applied a rubric to use to analyze the interviews. While this was quantitative data, it was not statistically significant due to the small sample size. However, we found the quantative data to be a very informative way to illustrate the qualitative results. As our goal was to identify the strengths and difficulties, not to present an in-depth analysis on a particular concept/skill, we found the quantitative data to be very informative. We also analyzed the videos qualitatively and describe the evidence for identifying a concept/skill as a strength or difficulty.

**TABLE 2.** Students interviewed in S22 and S24. The table shows their education level at the time they took the course, their level based on exam grades (High Proficiency (A) , Proficient (A-/B+), and Lower Proficiency (C), on exams), major, and whether or not they had taken quantum mechanics or linear algebra at the time of taking the course. For example, F21_1_a refers to a High Proficiency student interviewee from F21. The letters "a" and "b" distinguish individuals.

| | **Education level** | **Level based on exam grades** | **Major** | **Quantum** | **Linear Algebra** |
|---|---|---|---|---|---|
| **Fall 2021** | | | | | |
| F21_1_a | Senior | High Proficiency (1) | Physics | Yes | Yes |
| F21_1_b | Senior | High Proficiency (1) | Physics | Yes | Yes |
| F21_2_c | Senior | Proficient (2) | Physics | Yes | Yes |
| F21_3_d | Junior | Lower Proficiency (3) | Physics | Yes | Yes |
| **Fall 2023** | | | | | |
| F23_1_a | Senior | High Proficiency (1) | Computer Engineering | No | No |
| F23_2_b | Senior | Proficient (2) | Computer Engineering | No | Yes |
| F23_2_c | Senior | Proficient (2) | Computer Engineering | No | Yes |
| F23_3_d | Senior | Lower Proficiency (3) | Computer Engineering | No | No |

## A. Rubric Development and Analysis

Once the S24 interview questions had been chosen from the S22 interviews, we developed a rubric we could use to identify the skills and concepts evidenced in the interviewee answers. The rubric was designed to identify evidence of particular concepts or skills by students in answering the interview questions. The rubric used for each interview question is presented in Appendix B after the interview question. Each rater awarded a 1 or a 0 for each bullet point in the rubric for evidence of that skill or concept in the interview. Sometimes a concept or skill was not addressed in a particular interview because decisions were made by the interviewer to move on from a question when it seemed that continuation of the questioning sequence would be unproductive or based on available time to complete the interview. In these cases, a bullet point was awarded NA for not applicable.

There were four raters, two physics education research (PER) faculty and two computer science (CS) graduate students. The raters viewed the interviews from both S22 and S24, applying the rubric to each interview. We averaged the scores (1 or 0) across all raters for each concept/skill. Then we divided that average by the total possible points and compared the percentage for each interviewee for each concept/skill across semesters, attentive to the level of the students in each course. The quantitative results are shown in Figures 1 and 2. After each rater had coded all of the interviews, we used Fleiss Kappa [37, 38] as a measure of inter-rater reliability. Across the four raters, Fleiss Kappa was calculated to be 0.73, indicating substantial agreement across raters.

### B. Interviews

Qualitatively, we viewed the interview videos to identify common strengths and difficulties of students in understanding QC concepts and in the application of those concepts in theoretical analyses of quantum computing circuits and protocols. This project was not an in-depth analysis of the strategies used by different students on a particular concept/skill, but a broad study of evidence of common strengths and difficulties across many topics and possible evidence of the benefits and drawbacks of the mini-tutorials. However, we do describe the student responses and sometimes use direct quotes from representative interview segments as evidence to support our analysis.

The interviews were carried out the semester after the students took the course. So, the F21 students were interviewed in S22 and the F23 students were interviewed in S24. In S22, there were two interviews with each student volunteer. In S24 there was only time for one interview per student volunteer, so we chose a subset of the S22 questions to give us a comparison across semesters and to align as well as possible with some of the mini-tutorials. These included questions on normalization, probability, entangled states, single qubit gates, half-adder, phase kickback, and teleportation. Sample interview questions used in both S22 and S24 and the rubric employed are shown in Appendix B. In S22 there was a single interviewer. In S24 there were three different interviewers, but the same questions were asked by each interviewer. Both interviews were conducted a semester after the students took the course.

The interviews were semi-structured, so the interviewer could ask for more detail for clarification, if needed. If the student was stuck but might be able to progress with a little help, the interviewer would offer a hint, such as the matrix for a particular gate. The interviews were recorded using Zoom or some other recording platform.

## IV. RESULTS

We divided the strengths and difficulties into four broad categories: 1) Fundamental Quantum Mechanics (QM) Knowledge, 2) Multiple Representations, 3) Gates and 4) QC concepts and protocols. Within each category, we discuss the results from the interviews descriptively and discuss the quantitative results from the rubric. We found that the quantitative coding was an informative way to represent the strengths and difficulties. It is not meant to be interpreted as statistically significant but as a useful illustration that reflects the qualitative results across students and across semesters.

We present the results from the quantitative coding in Figures 1 and 2. In our qualitative discussion of the results of each category, we will refer to the students interviewed as labeled in Figures 1 and 2.

In Figure 1, we show the comparison of F21 and F23 students for each concept or skill as evidenced in the interview questions across levels. For each concept/skill, we averaged the scores (1 or 0) across all raters. We divided that average by the total possible points for that concept/skill and present it as the percentage of total possible points for each concept/skill based on the rubric. In some cases, the interviewer made a decision to move on from a question when it seemed that continuation of the questioning sequence would be unproductive or due to available time to complete the interview. If the question was not addressed in the interview, it appears as NA. Students at the same level are grouped together. So, for example, student a who took the course in F23 and is a level one student, would be labeled F23_1_a.

In Figure 2, we show the average of all students interviewed for each concept/skill for each semester and the overall average across both semesters. The averages do not include students assigned an NA because the question was not addressed in a particular interview.We have highlighted the averages in blue that might be misleading due to the number of students coded.

## A. Fundamental QM Knowledge

### *1. Normalization and Probability*

All of the High Proficiency students both semesters were coded at 100% or near 100% in conceptual understanding or skill application when answering probability and normalization questions in both cohorts. In the interviews, they answered correctly and quickly, without hesitation.

The F21 Proficient student also demonstrated a strong understanding of the concepts of normalization and probability, also answering quickly and correctly. The F23 Proficient students took longer to answer questions on normalization and probability correctly, but they took time to think through the questions carefully. F23_2_b, for example, first stated "..they all have to have an equivalent probability, which it looks like they do…" when asked about normalization. But the student did not recognize that the amplitudes as given in the problem would not normalize to one. Their discussion of probability, however, led them to return to rethink normalization. The student had written the two qubit state as the product of two one-qubit states and realized a square root of two was missing. They returned to the two-qubit state and recognized that each term needed to be multiplied by ½ for the state to be normalized so that the total probablity was equal to one.

The other F23 Proficient student, F23_2_c, initially recognized that the probability amplitudes squared needed to add to one for the state vector to be normalized and that you had to normalize the quantum state vector in order to determine the probabilities. However, they first wrote an equation for normalization of a single qubit state and struggled to apply the concepts to a multi-qubit state, as in the question asked.

| | | S24 | S22 | S22 | S24 | S24 | S22 | S24 | S22 |
|---|---|---|---|---|---|---|---|---|---|
| | | F23_1_a | F21_1_a | F21_1_b | F23_2_b | F23_2_c | F21_2_c | F23_3_d | F21_3_d |
| Fundamental QM knowledge | Normalization | 1 | 1 | 1 | 0.6 | 0.5 | 1 | 0.4 | 0.1 |
| | Probability | 1 | 0.9 | 1 | 1 | 0.1 | 0.9 | 0 | 0 |
| | Entanglement | 1 | NA | 0.5 | 0 | 0.4 | 0.9 | 0 | 0 |
| | | 1 | NA | 0 | 0 | 0.4 | 0.9 | 0 | 0 |
| Multiple Representations | Matrix operations | 1 | 1 | NA | 0 | 0 | NA | 0 | NA |
| | Dirac operations | 1 | 1 | 0.9 | 1 | 0 | NA | 1 | 0.5 |
| | Bloch Sphere | 0.7 | 0.4 | 0.3 | 0.4 | 0 | 0.3 | 0.7 | 0 |
| Gates | X | 1 | 1 | 1 | 1 | 1 | 1 | 0 | 0.1 |
| | | 0.9 | 1 | 1 | 1 | 0.7 | 1 | 0 | 0.9 |
| | S dagger | 1 | 1 | 0.4 | 1 | NA | 0 | NA | 0 |
| | H | 1 | 1 | 1 | 0.4 | 1 | 1 | 1 | 0.5 |
| | | 1 | 0.9 | 1 | 1 | 1 | 1 | 1 | 0 |
| | S | 1 | 1 | 0.4 | 1 | 0.1 | 0 | 0 | 0 |
| | Z | 1 | 1 | 0.4 | 1 | 0.4 | 0 | 0 | 0 |
| | CNOT | 1 | 1 | 1 | 1 | 0.8 | 1 | 1 | 1 |
| | | 1 | 1 | 0.7 | 1 | 0.5 | 0.7 | 0.2 | 0.6 |
| | Toffoli | 1 | 1 | 1 | 1 | 1 | 1 | 1 | 0.5 |
| QC concepts/protocols | Phase Kickback | 1 | 0.9 | NA | 0 | 0.7 | 0 | NA | NA |
| | Teleportation qualitative | 0.5 | 0.8 | 0.1 | 0.3 | 1 | 0.3 | 0.6 | 0.3 |
| | Teleportation quantitative | 0.72 | 0.8 | 0.5 | 0.3 | 0.2 | 0 | 0 | 0.3 |

**FIGURE 1.** Comparison of F21 and F23 students for each concept or skill as evidenced in the interview questions across levels. For each concept/skill, we averaged the scores (1 or 0) across all raters. We divided that average by the total possible points for that concept/skill and present it as the percentage of total possible points for each concept/skill based on the rubric. In some cases, the interviewer made a decision to move on from a question when it seemed that continuation of the questioning sequence would be unproductive or due to available time to complete the interview. If the question was not addressed in the interview, it appears as NA.

Students at the same level are grouped together. So, for example, student a who took the course in F23 and is a level one student, would be labeled F23_1_a.

| | | F23 average | F21 average | Average all interviewees |
|---|---|---|---|---|
| Fundamental QM knowledge | Normalization | 0.6 | 0.8 | 0.7 |
| | Probability | 0.5 | 0.7 | 0.6 |
| | Entanglement | 0.4 | 0.5 | 0.4 |
| | | 0.4 | 0.3 | 0.3 |
| Multiple Representations | Matrix operations | 0.3 | 1 | 0.4 |
| | Dirac operations | 0.8 | 0.8 | 0.8 |
| | Bloch Sphere | 0.5 | 0.3 | 0.4 |
| Gates | X | 0.8 | 0.8 | 0.8 |
| | | 0.7 | 1 | 0.8 |
| | S dagger | 1 | 0.4 | 0.6 |
| | H | 0.9 | 0.9 | 0.9 |
| | | 1 | 0.7 | 0.9 |
| | S | 0.5 | 0.4 | 0.4 |
| | Z | 0.6 | 0.4 | 0.5 |
| | CNOT | 0.9 | 1 | 1 |
| | | 0.7 | 0.7 | 0.7 |
| | Toffoli | 1 | 0.9 | 0.9 |
| QC concepts/protocols | Phase Kickback | 0.6 | 0.4 | 0.5 |
| | Teleportation qualitative | 0.6 | 0.4 | 0.5 |
| | Teleportation quantitative | 0.3 | 0.4 | 0.4 |

**FIGURE 2.** Average of all students interviewed for each concept/skill for each semester and the overall average across both semesters. The averages do not include students assigned an NA because the question was not addressed in a particular interview.We have highlighted the averages in blue that might be misleading due to the number of students coded.

The Lower Proficiency students either semester did not remember the concepts of normalization or probability very well. F23_3_d first stated that the sum of the probabilities would be zero but later realized that it needed to be one. However, the student did not remember how to normalize the state vector. In general, across all categories, the Lower Proficiency students often did not remember enough for us to identify difficulties.

For normalization and probability concepts, the difference in F21 and F23 Proficient and Lower Proficiency student answers may be due to whether or not they had taken quantum previously or concurrently with the QC course. However, there is clearly room for improvement on the mini-tutorials about probability and normalizaton based on the answers of Proficient students and inability to answer of the Lower Proficiency students in each cohort.

*2. Entanglement*

Except for student F23_1_a, the students did not demonstrate a solid understanding of the concept of entanglement in answering the interview questions. Some students, F21_1_b and F23_2_c, demonstrated understanding of the concept but had difficulty determining if the states given in the problem were entangled. They recognized that a two-qubit state is entangled if it cannot be written as the tensor product of two individual qubit states. They knew that they had developed an algebraic method in class to determine if a state was entangled or not. However, they had incorrectly memorized the result of the algebraic method. When they applied the incorreclty memorized result, they got an incorrect answer and they did not remember how to generate the algebraic method.

In problems worked in class in F21 and in the mini-tutorial in class in F23, students had been asked to come up with a method for determining if a two-qubit state was entangled or not. We will need to redesign the mini-tutorial on entanglement to focus more on using the definition of entanglement to devise a process to determine if a state is entangled or not, so that they focus on understanding the development of the process and not just the process result.

## B. Multiple representations

In the interviews, the students had been asked to represent a superposition state in three different representations: matrix, Dirac, and Bloch Sphere. The students were asked, if given a single qubit in either the |0> or |1> state, how they would put it in superposition. They were asked to do it using matrix notation, Bloch sphere representation, Dirac notation and to illustrate it using a circuit diagram.

Only one of the F21students answered correctly using matrix notation. Three F21 students (one at each level) answered correctly using Dirac notaton. One explicitly said he was not comfortable using matrix notation. The F23 students, except for F23_1_a, were also not comfortable with matrix notation in answering the interview question and preferred to work in Dirac notation. Student F23_2_b said, "So, like I said, I can work in Dirac notation all day, but linear algebra skills…". We observed in the interviews both years that the majority of the students preferred working with Dirac notation and applied it correctly.

F23 students demonstrated a better understanding of the Bloch sphere representation. Three out of four students could correctly identify the positions of |0> and |1> along the positive and negative directions along the z-axis and also drew axes for |+>, |->, |i>, and |-i>. However, they did not always use a righthand rule orientation. They were able to correctly identify a superposition state on the Bloch sphere and discussed how a Hadamard gate rotated |0> to |+> and |1> to |->. F23_1_a

was also able to identify the axis of rotation when a Hadamard gate was used to rotatate |0> into |+>. The fact that the F23 students demonstrated a better understanding of the Bloch sphere than the F21 students, indicates that the mini-tutorial may have been helpful. However, it will need some modification or some supplementation on matrix notation and, to a lesser extent, on Bloch sphere notation. While not commonly used in quantum mechanics, the Bloch sphere is an important representation in Quantum Computing.

### C. Gates

Most students in both semesters did well when applying the gates frequently used in class, the X, H, CNOT, and Tofolli gates. The students did not do well with the phase gates $S$ and $S^{\dagger}$, which are gates that were less frequently used in class. One would expect the Z gate to be understood as well as the X and H gates. We expect that it did not show up that way because in the question they were asked it was sandwiched between the $S$ and $S^{\dagger}$ gates with which students struggled.

In class, three methods were introduced for understanding different gates: matrix calculation, Dirac representation, and rotation around the Bloch sphere. The F23 interviewees received a mini-tutorial (Appendix A.2) that required them to determine the state of a qubit after a series of gates using all three methods, whereas the F21 interviewees did not receive this mini-tutorial.

Comparing students of similar performance levels, we found that for High Performance students, F23_1_a and F21_1_a mastered all gates, while F21_1_b incorrectly memorized Bloch sphere rotation methods. None of the students used the matrix method voluntarily.

For Proficient level students, some from each semester (F23_2_c and F21_2_c) struggled with the rotation gates, $S$ and $S^{\dagger}$, and were not successful in either Dirac or matrix notation. However, one Proficient student, F23_2_b, initially struggled with the output qubit from an S gate when the input qubit was $\frac{1}{\sqrt{2}}(|0\rangle+|1\rangle)$. While being prompted with the matrix for S gate, F23_2_b wrote down the output qubit as $\frac{1}{\sqrt{2}}(|0\rangle+i|1\rangle)$. This suggests that F23_2_b may have synchronized the representations of matrix and Dirac notation. Correspondingly, he did not use the matrix to calculate but analyzed what the matrix for an S gate could do to $|0\rangle$ and $|1\rangle$ respectively. The mini-tutorial about gates may have positively impacted F23_2_b's learning by engaging him in all three representations, facilitating their integration.

The Lower Proficiency students, F23_3_d and F21_3_d, did not demonstrate an operational understanding of the $S$ and $S^{\dagger}$gates in any of the notations, matrix, Dirac or Bloch sphere.

Overall, students in both classes were proficient in the application of the more frequently used (in class) single qubit gates and less proficient in the application of rotation gates, which were not used as often in class. There was some evidence (F32_2_b) that the mini-tutorials might have positively influenced the F23 students understanding and application of gates.

### D. QC Concepts/Protocols

#### *1. Phase kickback*

The concept of phase kickback was not addressed in all of the interviews. In the interviews where the concept was addressed, in F23, ⅔ of the students could identify phase kickback in a circuit and correctly discussed the concept. The other student did not remember the concept. In F21, in the interviews where the concept was addressed, one student could identify phase kickback in a circuit and remembered the concept and one student could not.

Students who could not identify phase kickback did not remember the concept and did not remember enough to identify difficulties. Students who did remember phase kickback demonstrated their understanding like F23_2_c, who said, “I mean phase kickback is when the target changes the control.” However, the student then applied the concept directly to a control qubit in the |+> state (changing it to |->) with the target in |-> (leaving it in |->) without writing out the target and control in terms of |0> and |1>.

Other demonstrations of a reasonable understanding of phase kickback came from F21_1_a and F23_1_a:

From F21_1_a: “...well that is fundamentally what phase kickback is when the target is flipping the control instead.”

From F23_1_a: “Since this one starts in a minus state instead of a plus state, you end up with these negative signs down here in your state, which normally you would see the CNOT gate being the qubit with the dot on it flipping the qubit with the plus on it. But in this case specifically, since your phase of your target qubit is it's in a minus state, then it kind of works the other way around if both of your bits are in this plus or minus state. Yeah, it flips the phase of the control qubit. So yeah, that's essentially what phase kickback is. It would flip the phase of the control bit when the phase of the target qubit is minus.”

*2. Quantum Teleportation*

In F21, only one student, F21_1_a, showed evidence of a qualitatively correct understanding of teleportation and was also able to demonstrate a reasonably good understanding of teleportation quantitatively, reproducing and discussing the circuit. The other High Proficiency student, F21_1_b, correctly discussed the no-cloning theorem and started to draw the circuit for teleportation, but just included Alice’s qubits in the circuit drawing. The student then started to carry out the math demonstrating the quantum state after Alice had entangled her unknown qubit with her qubit that was entangled with Bob’s qubit. The student started out correctly but could not factor out Alice’s qubits to clearly see how a measurement by Alice would determine the gate(s) Bob should apply. The student knew the correct answer but was having trouble demonstrating it mathematically.They did know how to interpret the four different possible answers of Alice’s measurement to determine which gate(s) Bob should apply to recreate Alice’s unknkown qubit for the given entangled state. So, the student did not discuss the process qualitatively in the interview, just immediately started drawing the circuit and writing out the state vector after Alice had entangled her two qubits.

The other F21 students, (Proficient, and Lower Proficiency), did not demonstrate sufficient qualitative descriptions. Their descriptions were vague, missing information or contained incorrect pieces of information.

Of the F23 students, F23_1_a and F23_2_b, demonstrated an incomplete qualitative understanding of quantum teleportation. Their answers were short and contained some correct and some incorrect information. F23_2_b, for example, said “Alice is trying to send a message to Bob and because she has one part of an entangled qubit, she’s able to entangle her message with that part of her entangled qubit. And then Bob can take a measurement and decode Alice’s measurement or Alice’s meassage.” The student had the correct general idea but did not recognize that it was Alice that made a measurement and sent those results classically to Bob so that Bob could reconstruct Alice's unknown qubit.

The other two F23 students, however, F23_2_c and F23_3_d, demonstrated a very good qualitative understanding of quantum teleportation. They discussed that qubits cannot be copied and how teleportation solves that problem. F22_3_d said "There's a big long no-cloning theorem that says why that can't happen. But one of the ways around that is you can use teleportation to kind of use the third bit to entangle the quantum information of the sender bit and the receiver bit…" The student continued with a correct description of the sender entangling the unknown qubit with their part of the entangled qubit from a third party, measuring both qubits and sending the results of that measurement to the receiver so the receiver could reconstruct the unknown qubit.

F23_2_c also gave a correct qualitative description of teleportation, discussing the no-cloning theorem, describing the story of "Alice" and "Bob", a third party who created an entangled state, Alice's measurements and Bob's recreation of Alice's unknown state: "So, my understanding is that it's sort of like you're technically cloning them, but because of the no-cloning theorem you can't technically clone a qubit. But so, like the whole story is like Alice wants to send Bob like qubits, and she's trying to come up with a way that you can do it over a very, very, very long distance because otherwise it would be too long. And so they use the help of Telemon. I think it's the name Telemon. And so he entangles the qubits. But yeah, so quantum teleportation is essentially like you entangle the qubits. So like Alice and Bob have different entangled qubits that they like they both get one part of that entangled qubit. So, when Alice goes through a set of like a circuit, and it gets two classical bits, and so she sends those two classical bits to Bob, which then can use those two classical bits to recreate Alice's qubits. So, you teleport them essentially. (Interviewer asks "What do you mean by recreate?") I mean when Alice sends Bob the like one and zero, right? Classical bits. Bob can then use like gates like not the Hadamard but like X, Z or Y, right, to recreate or essentially like copy, such as clone, what Alice had before that. So like if Alice sends him a 00 then he knows exactly what it is. If Alice sends him a 01 he has to put it through an X or a Z gate or something like that. So you kind of have like rules to follow to recreate it, but so that satisfied the no-cloning theorem, but it lets you quote, unquote teleport them."

One student, F23_1_a, demonstrated a good quantitative understanding of the circuit after receiving some help from the interviewer, constructing the circuit, representing the quantum state mathematically at each step and explaining the gates the receiver would apply to recreate the unknown state based on the sender's measurements.

Overall, the F23 students demonstrated a stronger qualitative understanding of quantum teleportation than the F21 students. This could be due to the mini-tutorial on quantum teleportation which required them to describe the technique qualitatively and to work through the circuit starting with a different Bell State than was discussed in class.

## V. DISCUSSION AND CONCLUSIONS

As the teaching of QC has expanded rapidly across many STEM fields and across many education levels from K-12 to graduate school, empirical research on students' understanding of and ability to apply fundamental QC concepts has only just begun. We have presented a comparison of two case studies of students' strenghts and difficulties when applying basic QC concepts and skills after taking an upper-level introduction to QC course. We have presented the results of interviews with student volunteers on topics that were covered early in the course they had taken the previous semester.

We have analyzed the strengths and difficulties of students in an interactive-engagement QC class taught with and without a set of mini-tutorials. We reported on the interviews with four students in each class in F21 and F23.

Across both classes, we found that most students demonstrated their ability to apply single-qubit and two-qubit gates correctly, except rotation gates. They also favored working in Dirac representation above matrix or Bloch sphere representations of qubits. However, the F23 class demonstrated a better understanding of the Bloch sphere representation than the F21 class. This could be due to the mini-tutorial that required them to use all three representations.

Many students in both classes demonstrated a reasonable understanding of and ability to apply the concepts of normalization and probability, although F21 students appeared to be more fluent in applying the concepts when answering questions. This could be be attributed to more practice with the concepts due to prior knowledge of quantum mechanics, as opposed to lack of impact of the mini-tutorials.

Even though the concept of phase kickback was not addressed in all of the interviews, the High Proficiency and Proficient students who were able to identify phase kickback in a circuit were able to describe the concept very well. This is an important concept in quantum computing and not one usually taught in quantum mechanics.

Overall, the F23 students demonstrated a stronger qualitative understanding of quantum teleportation than the F21 students. This could be due to the mini-tutorial on quantum teleportation which required them to describe the technique qualitatively and to work through the circuit starting with a different Bell State than was discussed in class.

In summation, common strengths were students' ability to apply single-qubit and two-qubit gates correctly, except rotation gates, and the use of Dirac notation. Common difficulties across both classes were matrix representation of qubits, use of rotation gates, and a quantitative (based on recalling and analyzing the circuit) understanding of teleportation.

Beyond identifying students' strengths and difficulties across both classes, we have identified mini-tutorials that may have been helpful to the F23 students and the need to modify or redesign some of the mini-tutorials. The entanglement and phase kickback mini-tutorials should be modified. Modifications are also needed for addressing rotation gates and the quantitative part of the teleportation mini-tutorial.

We have discussed the strengths and difficulties in each of the two classes and compared the strengths and difficulties of students across classes (our research questions 1 and 2.) For research question 3, we do not have strong evidence of the effect of the mini-tutorials. To address this, we will make some changes to our data collection in the future, including adding a set of pre- and post-tests to be administered as close to the beginning and end of the administration of the mini-tutorials as possible. We will also interview the students the semsester that they are taking the class, instead of the semester after they have taken the class.

Our limitations were the small sample size of both the classes and the number students who volunteered to be interviewed. Also, the interviews were the semester after the students took the course, not while they were taking the course.

## VI. ACKNOWLEDGMENTS

We would like to thank the National Science Foundation for their funding of DUE grant 2235464.

# APPENDIX A: SAMPLE MINI-TUTORIALS

## 1 Stern-Gerlach

1) It can be shown that the eigenvectors of the matrix representing a spin operator in an arbitrary direction $\hat{n}$ are:

$$|+>_n = \cos\left(\frac{\theta}{2}\right)|+> + \sin\left(\frac{\theta}{2}\right)e^{i\phi}\,|->$$
$$|->_n = \sin\left(\frac{\theta}{2}\right)|+> - \cos\left(\frac{\theta}{2}\right)e^{i\phi}\,|->$$

where

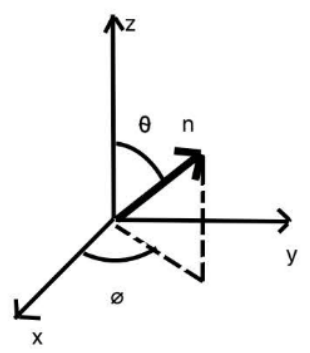

Consider a series of Stern-Gerlach analyzers as in the diagram below.

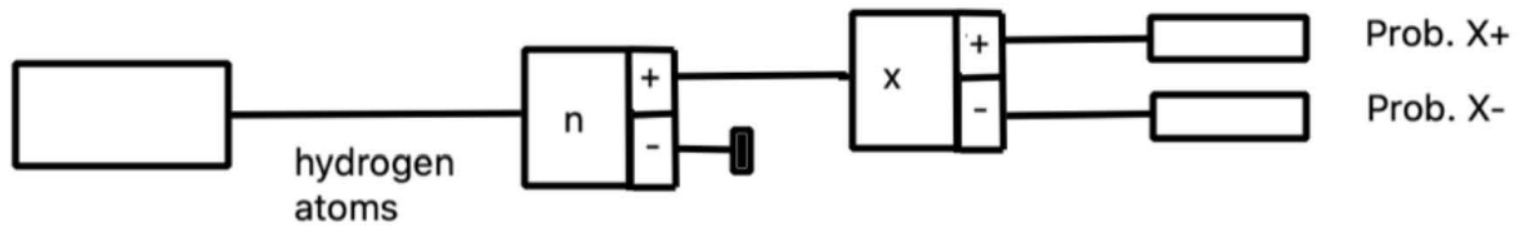

a) If the first Stern-Gerlach analyzer is oriented in a direction $\hat{n}$ determined by $\theta = \frac{\pi}{2}$ and $\phi = 0$,
   i) predict the probabilities of being found with spin up or spin down in the x-direction. Explain your reasoning.

   ii) determine the probabilities of being found with spin up or spin down in the x-direction. Show your process and discuss it with your neighbor.

b) If the first Stern-Gerlach analyzer is oriented in a direction $\hat{n}$ determined by $\theta = \frac{\pi}{3}$ and $\phi = \frac{\pi}{6}$,
   i) determine the probabilities of being found with spin up or spin down in the x-direction. Show your process.

   ii) Discuss what your result means.

## 2 Single Qubit Gates

1) Given the quantum circuit below,

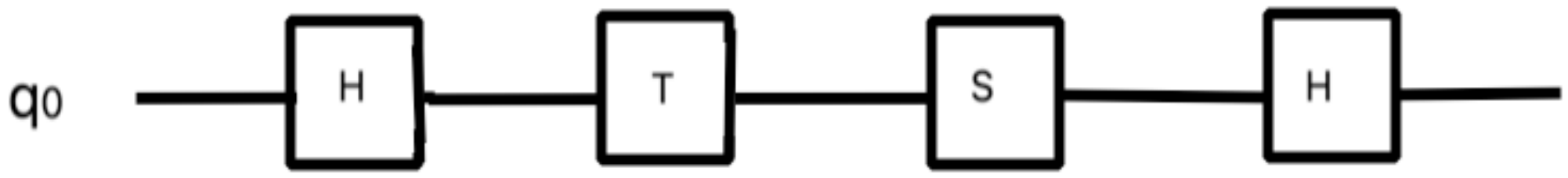

determine the state of the qubit after each gate and the possible results of a measurement in the computational basis after the last gate. Assume $q_0$ is initially set to |0>. Determine the state of the qubit after each gate three different ways, using each of the methods below:

a. Matrix method
b. Dirac method
c. Bloch sphere

## 3 Entangled States

1) Are the following states entangled? Find a method to determine if a state is entangled or not.

a) $\frac{1}{2}|11> + \frac{1}{\sqrt{2}}|10> + \frac{1}{2}|00>$

b) $1/2(|00\rangle + i|01\rangle - i|10\rangle + i|11\rangle)$

c) $1/\sqrt{2}(|\,00\rangle - |\,11\rangle)$

d) $1/2(|00\rangle + |01\rangle + i|10\rangle + |11\rangle)$

e) $1/2(|\,00\rangle + |\,01\rangle + 10\rangle + (-1)^n\ |\,11\rangle)$ for n even? For n odd?

## 4 Phase Kickback

1) Given the quantum circuits below, write out the qubit states for each circuit after the CNOT gate applies. $q_0$ and $q_1$ are initially set to |0>.

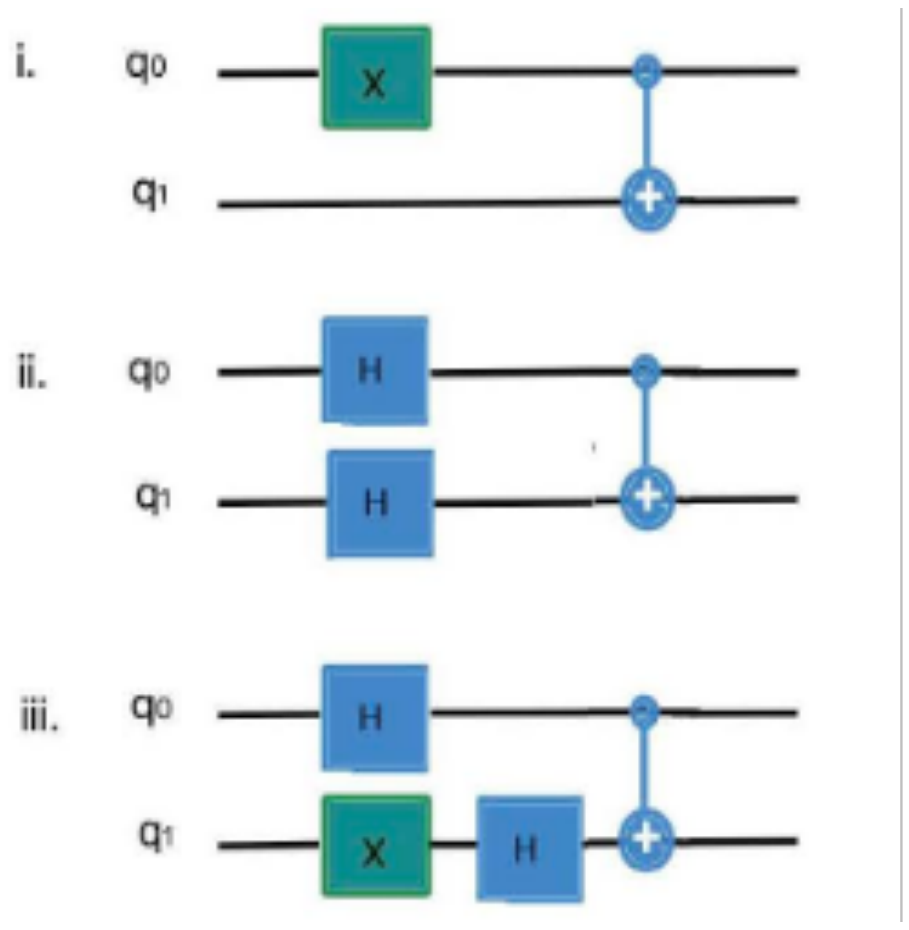

a) Show and explain how you arrived at the final states for each.
b) What can you say about the operation of the CNOT gate in each of these circuits?
c) What is the difference in the result in the CNOT gate in the second and third circuits? Explain.

## 5 Teleportation

1) Describe teleportation qualitatively.

2) In class, we worked through teleportation with the entangled qubits given to Alice and Bob in the $|\Phi_+ > = \frac{1}{\sqrt{2}}(|00 > + |11 >)$ state. Work through teleportation with the entangled qubits given to Alice and Bob in the $|\Psi_- > = \frac{1}{\sqrt{2}}(|01 > - |10 >)$ state. List Alice's possible measurement results and the operations Bob would need to apply to his qubit in each case.

## APPENDIX B: INTERVIEW QUESTIONS AND RUBRIC

After each interview question, the rubric is in red. Each bullet point was given a 1 or a 0 by the rater, indicating if the interview response contained evidence of that concept/skill or not.

Q1. Given the state $|q_1q_0\rangle = |00\rangle + |01\rangle + |10\rangle + |11\rangle$.

a. Is $|q_1q_0\rangle$ normalized? How were you able to tell?

b. If $|q1q0\rangle$ is not normalized, normalize $|q_1q_0\rangle$. Why is it important to normalize qubit states in quantum computing?

c. What is the probability of getting a zero when making a measurement of $q_0$? Why?

d. Suppose a measurement has been made and $q_0$ has been measured and found to be zero. What is now the probability of measuring the $q_1$ as zero? Why?

Normalization

- Understanding mathematically that the sum of coefficient squares is 1, $|<q_1q_0|q_1q_0>|$^2;
- Know how to normalize a state.

Probability

- Coefficient squares are the probabilities of states.
- Part c. Accurately calculate the probabilities of states.
- Part d. Accurately calculate the probabilities of states

Q2. Given the circuit below:

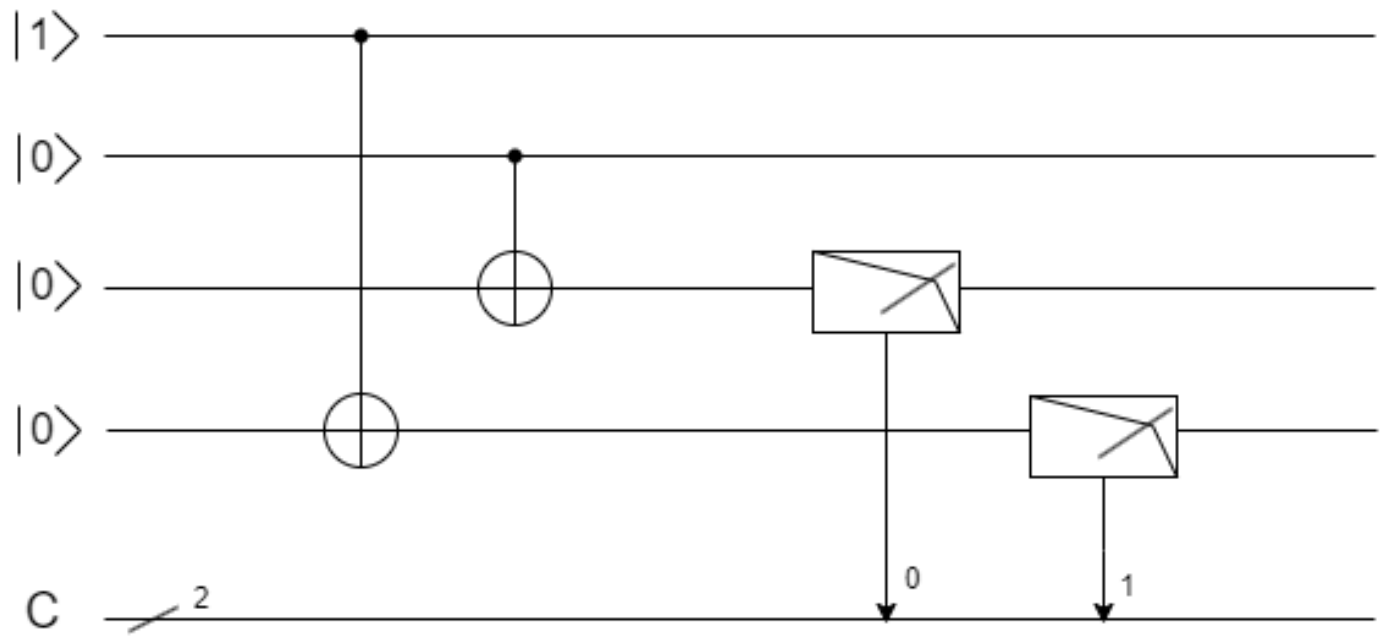

a. What outcome does the circuit give after measurement? Explain.
b. What does this circuit do? Show your work.

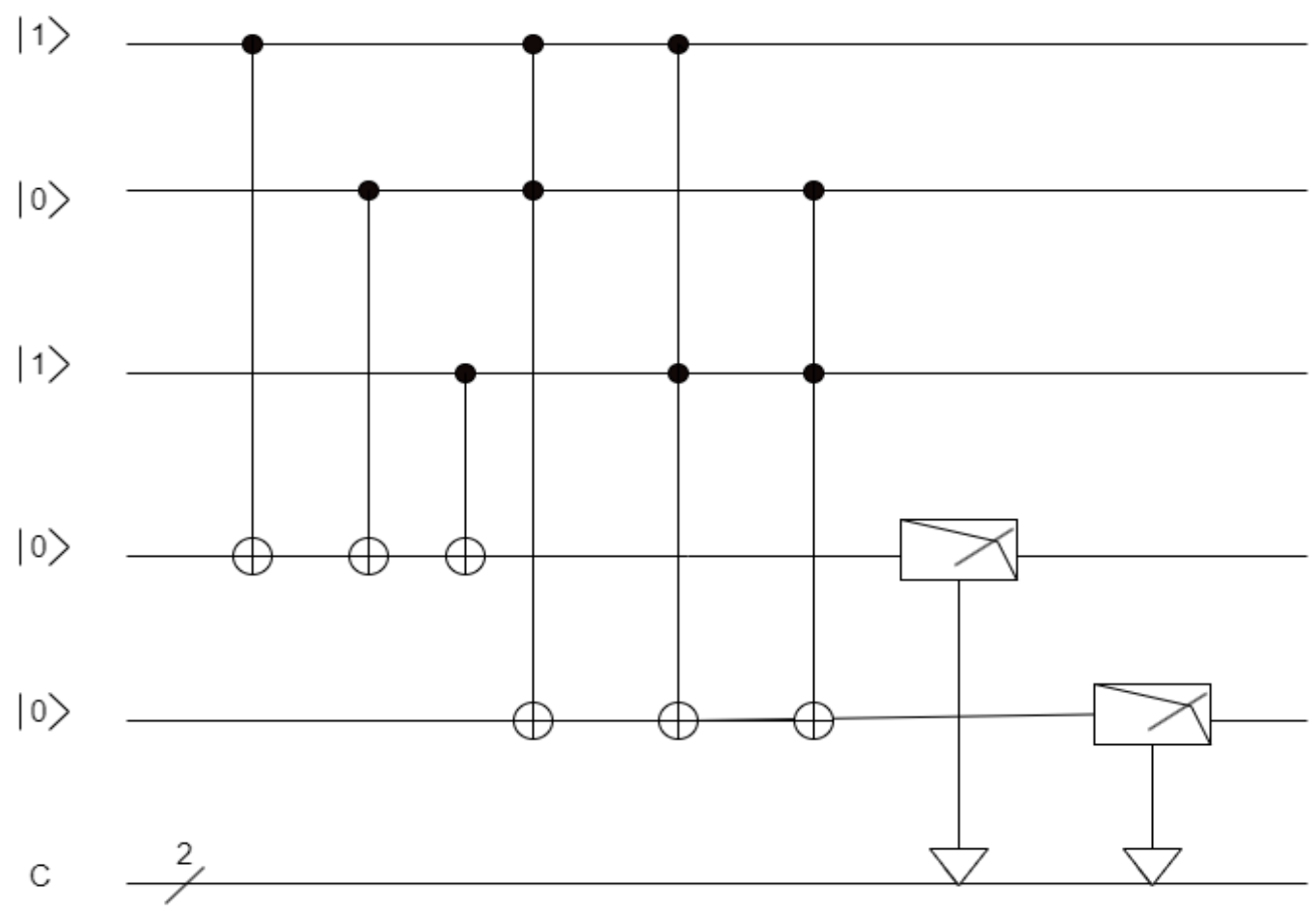

CNOT gate - a,b

- Circuit 1. When the control qubit is |1⟩, the target qubit flips.
- Circuit 2. When the control qubit is |1⟩, the target qubit flips.

Toffoli gate - b

- Circuit 2. When the control qubits are both |1⟩, the target qubit flips.

Measurement - a. b

- Circuit 1, The output should be written as 1 0, not as |1⟩ |0⟩
- Circuit 2, The output should be written as 1 0, not as |1⟩ |0⟩

Q3. Given a single qubit in either |0⟩ or |1⟩

a. how would you put it in a superposition state? Show your work in matrix notation .
b. Draw a circuit diagram of your response.
c. Show the state on a Bloch sphere
d. Write the state using Dirac notation

Matrix operation - a

- Use the matrix format of the H gate to present the transformation of |0⟩ or |1⟩ to a superposition.

Interpreting circuit diagrams - b

- Draw a circuit with an H gate with the correct input (i.e., |0⟩ or |1⟩) and output.

Bloch sphere - c

- Draw the Bloch sphere that accurately shows the positions of |0⟩ and |1⟩) along the positive and negative direction of the z axis.
- Draw the x and y axes of the Bloch sphere that follow the right-hand rule with the z axis.
- Identify the superstation state accurately on the Bloch sphere.

Diract notation - d

- Clearly write the input (i.e., |0⟩ or |1⟩) and output (i.e., |+⟩ or |-⟩) in Dirac notations

Superposition - a,b,c,d

- Understand that $|+\rangle$ and $|-\rangle$ are superposition states

Q4. What would be the final state(s) in the circuits below after the operations of the quantum gates on the initial state?

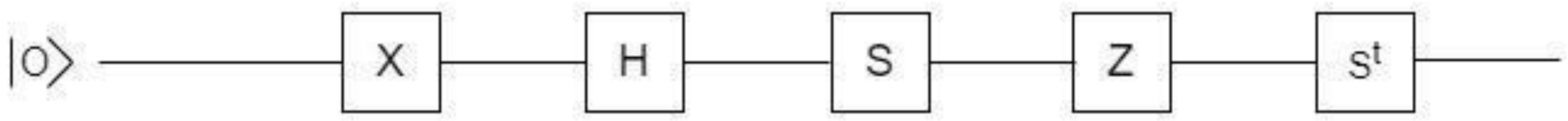

X gate

- Accurately identify the output of the X gate as $|1\rangle$

H gate

- Accurately identify the output of the H gate from the input (expected: $|-\rangle$)

S gate

- Accurately identify the output of the S gate from the input (expected: $\frac{1}{\sqrt{2}}(|0\rangle - i|1\rangle)$)
- Demonstrate reasonable justification through matrix operation or Bloch sphere rotation.

Z gate

- Accurately identify the output of the Z gate from the input (expected: $\frac{1}{\sqrt{2}}(|0\rangle + i|1\rangle)$)
- Demonstrate reasonable justification through matrix operation or Bloch sphere rotation.

S dagger

- Accurately identify the output of the S dagger gate from the input (expected: $|+\rangle$)
- Demonstrate reasonable justification through matrix operation or Bloch sphere rotation.

Q5. Write out the state of the multiple qubits at every stage in the circuits below.

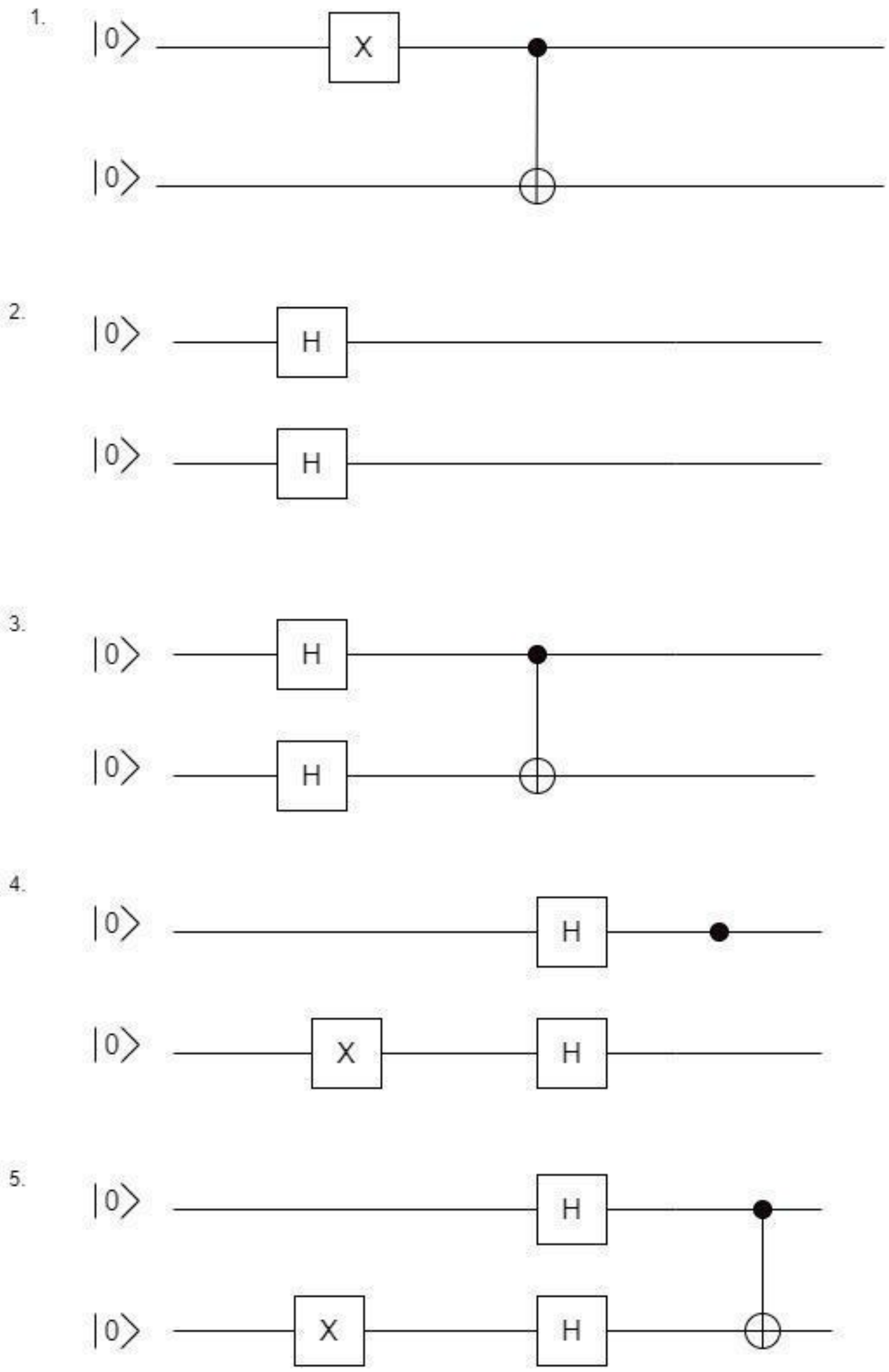

X-gate

- Circuit 1, Recognize when an X-gate is a applied to a single qubit it flips the qubit, taking a |0> to a |1> or a |1> to a zero.
- Circuit 4, Recognize when an X-gate is a applied to a single qubit it flips the qubit, taking a |0> to a |1> or a |1> to a zero.
- Circuit 5, Recognize when an X-gate is a applied to a single qubit it flips the qubit, taking a |0> to a |1> or a |1> to a zero.

CNOT gate

- Circuit 1, When the control qubit is |0> nothing changes. When the control qubit is |1>, the target qubit flips.
- Circuit 3, CNOT applied to a superposition state: The control and target qubits should be expressed in the computational basis and written as a two qubit state, keeping track of the ordering of q0 and q1 (control and target). Apply the gate term by term to the two-qubit state.

- Circuit 5, CNOT applied to a superposition state: The control and target qubits should be expressed in the computational basis and written as a two qubit state, keeping track of the ordering of q0 and q1 (control and target). Apply the gate term by term to the two-qubit state.

Hadamard gate

- Circuit 2, The Hadamard gate takes a |0> to a |+> and a |1> to a |->, putting the qubit in a superposition state. (Or rotating the qubit into the x-basis.) The student should be able to write out |+> = (|0> + |1>)/sqrt(2) and |-> = (|0> - |1>)/sqrt(2).
- Circuit 3, The Hadamard gate takes a |0> to a |+> and a |1> to a |->, putting the qubit in a superposition state. (Or rotating the qubit into the x-basis.) The student should be able to write out |+> = (|0> + |1>)/sqrt(2) and |-> = (|0> - |1>)/sqrt(2).
- Circuit 4, The Hadamard gate takes a |0> to a |+> and a |1> to a |->, putting the qubit in a superposition state. (Or rotating the qubit into the x-basis.) The student should be able to write out |+> = (|0> + |1>)/sqrt(2) and |-> = (|0> - |1>)/sqrt(2).
- Circuit 5, The Hadamard gate takes a |0> to a |+> and a |1> to a |->, putting the qubit in a superposition state. (Or rotating the qubit into the x-basis.) The student should be able to write out |+> = (|0> + |1>)/sqrt(2) and |-> = (|0> - |1>)/sqrt(2).

a. What can you say about the operation of the CNOT gate for the first, third and fifth quantum circuits?
b. What do you understand as phase kickback?

Phase kickback -a.b.

- Recognize that in 1. and 3. the state is unchanged and in 5., the target remains the same but the control qubit changes phase. (5. is an example of phase kickback.)
- Recognize (by showing wavefunction in computational basis and applying gates) that the initial state is |->|+> and the final state is |->|->. The target qubit remains the same and the control qubit has changed.

Q6. Identify the entangled states in the list below and explain why. Show your work.

a. $\frac{1}{2}(|00\rangle + |01\rangle + |10\rangle + |11\rangle)$

Entanglement

- Accurately identify this state as not entangled.
- Provide reasonable justifications, such as recognizing that the state can be factored into |+>|+>, or using the formula, or being able to generate a rule to determine if a state is entangled or not.

b. $\frac{1}{\sqrt{2}}(|{++}\rangle - |{--}\rangle)$

- Accurately identify this state as entangled.

- Provide reasonable justifications, such as recognizing that the state can be factored into |+>|+>, or using the formula, or being able to generate a rule to determine if a state is entangled or not.

Q7. What do you understand about quantum teleportation?

a. Could you explain quantum teleportation using the story of Alice and Bob?

Teleportation qualitative

- Alice has an unknown qubit that she wants to send to Bob but she cannot clone it (no-cloning theorem), so she needs another method.
- A third party (called Telemon in Qiskit) creates an entangled pair of qubits and sends one to Alice and one to Bob.
- Alice entangles her unknown qubit with the qubit from Telemon and measures both qubits. She sends the results of the measurements *classically* to Bob.
- Depending on the classical measurement results, Bob applies a particular gate to his qubit, producing Alice's unknown qubit. (Neither Alice or Bob know the state of the unknown qubit, it has simply been teleported to Bob.)

b. Can you draw the circuit diagram for such transfer of information and explain the wave function after each operation in the circuit?

Teleportation quantitative

- Correctly render the circuit diagram, including the creation of an entangled state by Telemon, Alice's application of CNOT and H, Alice's measurement.
- Accurately write the state after the creation of an entangled state by Telemon.
- Accurately write the state after Alice's application of CNOT and H.
- Accurately illustrate Bob's options of using possible gates based on Alice's measurement.

**References**

1. B. Cervantes, G. Pasante, B. Wilcox, amd S. Pollack, “An Overview of Quantum Information Science Courses at US Institutions,” 2021 PERC Proceedings, virtual conference, 93 (2021).
2. S. SeVore and C. Singh, “Interactive learning tutorial on quantum key distribution,” Phys. Rev. Phys. Ed. Rev. 16, 010126 (2020).
3. A. Kohnle, C. Bailey, A. Campbell, N. Korolkova, and M. Paetkau, “Enhancing student learning of two-level quantum systems with interactive simulations,” Am. J. Phys. **83** (6), 560, (2015).
4. A. Kohnle and E. Deffebach, “Investigating student understanding of quantum entanglement,” 2015 PERC Proceedings, College Park, MD, 171 (2015).
5. G. Pasante, P. Emigh and P. Shaffer, “Student ability to distinguish between superposition states and mixed states in quantum mechanics,” Phys. Rev. Special Topics Phys. Ed. Res., **11**, 020135 (2015).
6. S. Satanassi, E. Ercolessi and O. Levrini, “Designing and implementing materials on quantum computing for secondary school students: The case of teleportation,” Phys. Rev.Phys. Ed. Res. **18**, 010122 (2020).
7. A. Kohnle and A. Rizzoli, “Interactive simulations for quantum key distribution,” Eur. J. Phys. 38, 035403 (2017).
8. J. Meyer, G. Passante, S. Pollock, M. Vignal, and B. Wilcox, “Investigating students’ strategies for interpreting quantum states in an upper-division quantum computing course,” 2021 PERC Proceedings, virtual conference, 289 (2021).
9. J. Meyer, G. Passante, S. Pollock and B. Wilcox, “Investigating student interpretations of the differences between classical and quantum computers: Are quantum computers just analog classical computers?” 2022 PERC Proceedings, Grand Rapids, MI, 317 (2022).
10. P. Hu, Y.Li, R. Mong, and C. Singh, “Student understanding of the Bloch sphere,” Eur. J. Phys. **45**, 025705 (2024).
11. Hu, P., Li, Y., & Singh, C. (2024). Investigating and improving student understanding of the basics of quantum computing. Physical Review Physics Education Research, 20(2), 020108. https://doi.org/10.1103/PhysRevPhysEducRes.20.020108
12. T. Kushimo and B. Thacker, "Investigating Students' Strengths and Difficulties in Quantum Computing," 2023 IEEE International Conference on Quantum Computing and Engineering (QCE), Bellevue, WA, USA, 2023, pp. 33-39, doi: 10.1109/QCE57702.2023.202 (2023).
13. C. Singh, M. Belloni and W. Christian. "Improving students’ understanding of quantum mechanics." *Physics Today* 59.8 (2006): 43-49.
14. W. Marshman and C. Singh. "Framework for understanding the patterns of student difficulties in quantum mechanics." *Physical Review Special Topics-Physics Education Research* 11.2 (2015): 020119.
15. C. Singh and E. Marshman. "Review of student difficulties in upper-level quantum mechanics." *Physical Review Special Topics—Physics Education Research*11.2 (2015): 020117.
16. L. Bartlett and F. Vavrus, “Comparative case studies: An innovative approach,” Nordic Journal of Comparative and International Education (NJCIE), 1(1) (2017).

17. McDermott, L.C., Shaffer, P.S., *Tutorials in Introductory Physics* (Pearson, Boston, MA, 2013).
18. Bada, S. O., "Constructivism Learning Theory: A Paradigm for Teaching and Learning," IOSR-JRME, **5**, 6, 66-70, (2015).
19. Constructivist Learning Theory, https://elmlearning.com/hub/learning-theories/constructivism/, 05/08/2025.
20. Bransford, J. D., Brown, A. L., & Cocking, R. R. How people learn (Vol. 11). Washington, DC: National academy press (2000).
21. David H. McIntyre, *Quantum Mechanics: A Paradigms Approach*, (1st ed.). (Pearson, Boston, MA, 2012)
22. *Qiskit* textbook by IBM (legacy version), https://github.com/Qiskit/textbook/tree/main/notebooks/ch-demos#, 06/23/24.
23. Michael A. Nielson and Isaac L. Chuang, *Quantum Computation and Quantum Information* (2nd ed.). (Cambridge University Press, Cambridge, UK, 2010).
24. Thomas G. Wang, *Introduction to Classical and Quantum Computing.* (Rooted Grove, Omaha, Nebraska, 2022)
25. Reinhold Blumel, *Foundations of Quantum Mechanics: From Photons to Quantum Computers* (1st ed.), (Jones & Bartlett Learning, Burlington, MA, 2009).
26. Venksteswaran Kasirajan, *Fundamentals of Quantum Computing.* (Springer, New York, NY, 2021).
27. Ray LaPierre, *Introduction to Quantum Computing,*. (Springer, New York, NY, 2021).
28. Wolfgang Scherer, *Mathematics of Quantum Computing.* (Springer, New York, NY, 2019).
29. Bernard Zygelman, *A First Introduction to Quantum Computing and Information.* (Springer, New York, NY, 2018).
30. Franklin De Lima Marquezino, Renato Portugal and Carlile Lavor, *A Primer on Quantum*
31. *Computing,* (1st ed.). (Springer, New York, NY, 2019).
32. Jack D. Hidary, *Quantum Computing an Applied Approach.* (Springer, New York, NY, 2019).
33. Robert S. Suter, *Dancing with Qubits: How quantum computing works and how it can*
34. *change the world.* (Packt Publishing, Birmingham, UK, 2019).
35. Robert S. Suter (2021). *Dancing with Python: Learn to code with Python and Quantum Computing*. (Packt Publishing, Birmingham, UK, 2021).
36. J. Clement, Analysis of clinical interviews: Foundations and model viability. In *Handbook of research design in mathematics and science education* (pp. 547-589). (Routledge, Oxfordshire, UK, 2012).
37. J. Cohen, A coefficient of agreement for nominal scales. *Educational and psychological measurement*, **20**(1), 37 (1960).
38. J. L. Fleiss, "Measuring nominal scale agreement among many raters," *Psychological bulletin*, **76**(5), 378 (1971).